\documentclass[twocolumn,notitlepage,footinbib,amsmath,amssymb,amsfonts]{revtex4-1}

\usepackage{graphicx}
\usepackage{multirow}
\usepackage{color}
\usepackage{enumerate}
\usepackage[normalem]{ulem}
\usepackage{ytableau}
\usepackage[hypertexnames=false]{hyperref}

\newcommand{\mb}[1]{\ensuremath{\mathbf{#1}}}
\newcommand{\mc}[1]{\ensuremath{\mathcal{#1}}}
\newcommand{\mr}[1]{\ensuremath{\mathrm{#1}}}

\newcommand{\tr}{\ensuremath{\mathrm{Tr}}}

\newcommand{\im}{\ensuremath{\mathrm{Im}}}
\newcommand{\la}{\ensuremath{\langle}}
\newcommand{\ra}{\ensuremath{\rangle}}

\newcommand{\ket}[1]{\ensuremath{| #1 \rangle}}
\newcommand{\prj}[1]{\ensuremath{| #1 \rangle \langle #1 |}}

\newcommand{\Eq}[1]{Eq.~\eqref{#1}}

\newcommand{\Fig}[1]{Fig.~\ref{#1}}
\newcommand{\FIG}[1]{Figure~\ref{#1}}

\newcommand{\Ref}[1]{Ref.~\cite{#1}}

\newcommand{\Tbl}[1]{Table~\ref{#1}}
\newcommand{\Sec}[1]{Sec.~\ref{#1}}

\allowdisplaybreaks

\begin{document}

\title{Generalized Schrieffer-Wolff transformation of multi-flavor Hubbard models}
\author{Seung-Sup B. Lee}
\author{Jan von Delft}
\author{Andreas Weichselbaum}
\affiliation{Physics Department, Arnold Sommerfeld Center for Theoretical Physics and Center for NanoScience, Ludwig-Maximilians-Universit\"{a}t M\"{u}nchen, Theresienstra{\ss}e 37, 80333 M\"{u}nchen, Germany}
\date{\today}
\begin{abstract}
We give a self-contained derivation of the low-energy effective interactions of the SU($N$)
Hubbard model, a multiflavor generalization of the one-band Hubbard model, by using a generalized Schrieffer-Wolff transformation (SWT).
The effective interaction of doublons and holons, which has been largely ignored in previous SWT studies (e.g., the $t$-$J$ model), leads to distinct peaks in the local density of states. As shown by Lee {\it et al.}
\href{https://doi.org/10.1103/PhysRevLett.119.236402}{[Phys.~Rev.~Lett. {\bf 119}, 236402 (2017)]},
this underlying effective doublon-holon interaction explains the numerical observation of the subpeaks at the inner edges of the Hubbard bands in the metallic phase close to the Mott transition.
\end{abstract}

\maketitle

\section{Introduction}

The Schrieffer-Wolff transformation (SWT)~\cite{Schrieffer1966} 
was introduced in the 1960's to derive the Kondo model as an effective low-energy Hamiltonian for the single-impurity Anderson model.
Since then, variations and generalizations of its basic idea have been used in numerous contexts in condensed matter physics,
to map many-body Hamiltonians onto simpler effective Hamiltonians describing the effective degrees of freedom accessible at low excitation energies.

One paradigmatic application concerns
the one-band Hubbard model~\cite{Harris1967,*Chao1977,*MacDonald1988,*Eskes1994,*Eskes1994a}.
From the nearest-neighbour particle hopping and the on-site Coulomb interaction in the Hubbard model,
the SWT yields, up to leading order, several different types of
effective interactions:
an exchange interaction between nearest-neighbour spins,
an interaction between doublon and holon on nearest neighbours,
and three-site terms involving next-nearest neighbours.
Here we define doublon (holon) as the local excitation having
one more (less) particle than the average occupation number,
which naturally generalizes to the multi-flavor case discussed below.

When the $t$-$J$ model~\cite{Harris1967,*Chao1977,*MacDonald1988,*Eskes1994,*Eskes1994a}, an effective description of doped Mott insulators,
is derived from the Hubbard model,
the effective interactions are projected onto charge sectors
which select a specific integer filling,
and one allows charge fluctuations into
either the doublon or the holon sector,
but not both.
This inequivalent treatment for doublon and holon
is appropriate for the description of doped Mott insulators
in which doublon and holon hardly coexist,
since the excitation energy of doublon (holon) is much higher than that of holon (doublon)
in the hole-doped (particle-doped) case~\cite{Lee2017b}.
Thus the doublon-holon interaction drops out,
and three-site terms are neglected as an additional approximation.
As a result, only the exchange interaction, 
among all the effective interactions mentioned above, 
survives, yielding 
the $t$-$J$ model.

Unlike the exchange interaction, which is at the heart of quantum magnetism,
the role of the doublon-holon interaction has not received much attention so far.
However, the doublon-holon and exchange interactions have strengths comparable in magnitude.
Therefore one may suspect
that the doublon-holon interaction can lead to
measurable phenomena of its own,
especially when the coexistence of doublon and holon is substantial.

In a related paper~\cite{Lee2017}, we have identified a situation where this is indeed the case:
the doublon-holon interaction is responsible for
previously unexplained subpeaks at the inner edges of the Hubbard bands of the local spectral function, i.e., the local density of states.
Many dynamical mean-field theory (DMFT) studies~\cite{Zhang1993,Karski2005,*Karski2008,Ganahl2014,*Ganahl2015,Wolf2014,Zitko2009,Granath2012,Lu2014}
have observed such subpeaks emerge in the metallic phase close to the Mott transition.
Though the other spectral features in the spectral
function, i.e., the quasiparticle peak and the Hubbard
bands, are well understood,
the physical origin of the subpeaks had remained unknown.
In \Ref{Lee2017} we have proposed an explanation for their origin:
they arise from doublon-holon correlations.
To demonstrate this, we numerically computed the correlation functions of doublons and holons,
finding peak structures that indeed correlate with those of the subpeaks in the local density of  states.
Moreover, we also argued that the generic features of these subpeaks can be understood by using an effective low-energy model derived by a generalized SWT, and treated by a mean-field decoupling scheme.

In this work, we provide a concise self-contained derivation of the generalized SWT employed in \Ref{Lee2017}.
First, we derive the low-energy effective Hamiltonian of the $\mr{SU}(N)$ Hubbard models
by employing a generalized Schrieffer-Wolff transformation (SWT)~\cite{Bukov2016,Bukov2015}
inspired by a high-frequency expansion~\cite{Rahav2003} in the Floquet theory.
Then we study the correlation functions of doublons and holons, focusing on the intermediate energy scale
which lies between the larger energy scale associated with the Hubbard bands and the smaller scale with the quasiparticle peak.
By adopting a mean-field decoupling scheme, we briefly analyze the peak structure
in the local spectral functions related to the doublon-holon dynamics.
Finally, we argue that the peak structure 
becomes more pronounced with increasing number $N$ of particle flavors,
since the doublon-holon interaction acts on larger Hilbert space.
This is consistent with DMFT results from \Ref{Lee2017}.

\section{Low-energy Hamiltonian}
\label{sec:SWT}

\subsection{Generalized SWT}
\label{sec:Heff_SUN}

We consider the $\mr{SU}(N)$ Hubbard model,
which is the simplest multi-flavor generalization
of the one-band Hubbard model for $N=2$ spinful flavors.
The Hamiltonian for the $\mr{SU}(N)$ Hubbard model,
\begin{equation}
H = \underbrace{\frac{U}{2} \sum_i (\hat{n}_i - \bar{n})^2}_{\equiv H_U}
\underbrace{-\mu \sum_i \hat{n}_i}_{\equiv H_\mu}
+ \underbrace{\sum_{i,j; \nu} v_{ij} c_{i\nu}^\dagger c_{j\nu}}_{\equiv H_v}
\text{.}
\label{eq:Hubbard}
\end{equation}
describes $N$ symmetric flavors of fermions on a lattice, with
Coulomb interaction strength $U$, chemical potential $\mu$, and hopping amplitude $v$.
Consequently, the Hamiltonian has $\mr{SU}(N)$ flavor
symmetry.
In this paper, we focus on the cases in which the average occupation is integer, i.e., $\langle \hat{n}_i \rangle = \bar{n} \in \mathbb{N}_0$.
In \Eq{eq:Hubbard},
$\bar{n}$ is the average (integer) occupation of interest,
and $c_{i\nu}$ annihilates a particle of flavor $\nu = 1, \ldots, N$ at site $i$.
The Coulomb interaction $H_U$ yields finite energy when the particle number
$\hat{n}_i \equiv \sum_\nu c_{i\nu}^\dagger c_{i\nu}$ at each site $i$ deviates from $\bar{n}$.
By preferring finite occupation $\bar{n}$, this term
contains a significant portion of the chemical potential.
With $\tfrac{U}{2} (\hat{n}_i - \bar{n})^2 = \tfrac{U}{2} \hat{n}_i (\hat{n}_i - 1) + U (\tfrac{1}{2} - \bar{n}) \hat{n}_i + \text{const.}$,
the first term on the r.h.s.~is the ``bare'' Coulomb interaction,
whereas the second term represents an offset
$U (\bar{n} - \tfrac{1}{2})$ to the chemical potential.

The term $H_\mu$ describes an additional
fine tuning of the chemical potential to ensure
$\la \hat{n}_i \ra = \bar{n}$.
For example, particle-hole symmetry is given by $\bar{n} = N/2$ and $\mu = 0$,
resulting in half filling $\la \hat{n}_i \ra = N / 2$.
Otherwise for $\la \hat{n}_i \ra = \bar{n} \neq N/2$,
one typically has $\mu\neq 0$.
Since, by construction, the term $H_U$ contains the largest contribution to
the total chemical potential, 
we typically have $|\mu| \ll U$.

Finally, the kinetic term $H_v$ represents a hopping between nearest
neighbours with real hopping amplitude $v$,
by defining $v_{ij} = v_{ji} = v$ when sites $i$ and $j$ are nearest
neighbours and $v_{ij} = 0$, otherwise.

\subsubsection{Projected operators}

The first step of the generalized SWT is to 
identify the dominant high-energy term to be integrated out.
In \Eq{eq:Hubbard}, this is the Coulomb interaction term $H_U$
at energy scale $U$.
Given that the Coulomb interaction is solely sensitive to
the local occupation number of a specific site, it will be important
to meticulously keep track of the local site occupation when
considering individual hopping events as part of the kinetic energy
term in the Hamiltonian.

This requires to introduce the projectors
$P_{i,n}$ onto the subspace where a site $i$ has $n$ particles,
with the completeness relation
\begin{eqnarray}
     \sum_{n=0}^{N} P_{i,n} = 1.
\label{eq:Pin}
\end{eqnarray}
Then we can decompose a particle operator $c_{i\nu}^\dagger$,
\begin{equation}
c_{i\nu}^\dagger = \sum_{n=1}^{N}
     \underbrace{P_{i,n} c_{i\nu}^\dagger}_{\equiv \tilde{c}_{i\nu;n}^\dagger}
     \equiv \sum_n  \tilde{c}_{i\nu;n}^\dagger \, .
\label{eq:cin}
\end{equation}
Here, by definition, 
$\tilde{c}^\dagger_{i\nu;n}$ creates
a particle at site $i$ leading to a final occupation of $n$ particles.
Conversely, $\tilde{c}_{i\nu;n}$ destroys a particle
starting from an initial site occupation of $n$ particles.
Note that $\tilde{c}_{i\nu;n} = c_{i\nu} P_{i,n} = P_{i,n-1} c_{i\nu}$
are projected operators (also called Hubbard operators), and hence
no longer satisfy canonical fermionic
anticommutation relations. We will use the tilde as a
reminder of this fact throughout.
As already indicated on the r.h.s.\ of \Eq{eq:cin},
the range of $n$ is $[1, N]$,
unless specified otherwise.

For sufficiently large $U$,
large charge fluctuations $|n_i - \bar{n}| > 1$
will be suppressed in low-energy subspace due to
large Coulomb energy cost.
(Here $n_i$ stands for the eigenvalue of an operator $\hat{n}_i$.)
So we distinguish doublon and holon operators, 
\begin{equation}
d_{i\nu}^\dagger \equiv \tilde{c}_{i\nu;\bar{n}+1}^\dagger , \qquad
   h_{i\nu}^\dagger \equiv \tilde{c}_{i\nu;\bar{n}},
\label{eq:dh:gen}
\end{equation}
from the other projected operators $\tilde{c}_{i\nu;n}$.
They describe the more relevant excitations in low-energy subspace.
In the following derivation, however,
we will consider all possible contributions to the effective Hamiltonian in an unbiased way.
Whether each contribution is relevant or not will be determined by the nature of the phase we study,
after all possible contributions are collected first;
see \Sec{sec:SWT_interpretation} for more details.

\subsubsection{Rotating frame}

In a second step, the generalized SWT considers a rotating frame
whose time evolution is generated by the term $H_U$ with the largest
energy scale $U$.
A state $\ket{\psi (t)}$ in the lab frame is transformed to
$\ket{\psi_\mr{rot} (t)} = e^{i H_U t} \ket{\psi (t)}$ in the rotating frame.
This state evolves in time via the Schr\"{o}dinger equation $i \tfrac{d}{dt} \ket{\psi_\mr{rot}} = H_\mr{rot} (t) \ket{\psi_\mr{rot}}$
in which the rotating frame Hamiltonian $H_\mr{rot}$ is related
to the lab frame as
\begin{equation}
H_\mr{rot} \equiv -H_U + e^{i H_U t} H e^{-i H_U t} . \label{eq:Hrot}
\end{equation}%
In evaluating the term $e^{i H_U t} H e^{-i H_U t}$, 
the structure of the Baker-Campbell-Hausdorff formula, 
$e^X Y e^{-X} = Y + [X,Y] + \tfrac{1}{2!} [X,[X,Y]] + \tfrac{1}{3!} [X,[X,[X,Y]]] + \ldots$,
suggests that it is convenient to decompose the hopping term as
\begin{equation}
H_v = \sum_{i,j;\nu} v_{ij} c_{i\nu}^\dagger c_{j\nu}
\overset{(\ref{eq:cin})}{=}
\sum_{n,n'}
\underbrace{\sum_{i,j;\nu} v_{ij} \tilde{c}_{i\nu;n}^\dagger \tilde{c}_{j\nu;n'}}_{\equiv H_{v;nn'}} ,
\label{eq:Hvsplit}
\end{equation}
where the constrained hopping terms $H_{v;nn'}$ only include hoping between sites with fixed initial occupations. 
Then the corresponding energy cost for
this hopping process is described by the special structure
of the commutator
\begin{eqnarray}
    [ H_U, H_{v;nn'} ] = (n-n')U \cdot H_{v;nn'} ,
\label{eq:HU:comm}
\end{eqnarray}
where, importantly, $H_{v;nn'}$ again occurs intact on the RHS.
The prefactor $(n-n')U$ on the r.h.s.\ is the cost of Coulomb energy
to arrive at the final charge configuration after acting
with $H_{v;nn'}$ on the initial charge configuration.
Consider then, e.g., the hopping process
$\tilde{c}_{i\nu;n}^\dagger \tilde{c}_{j\nu;n'}$ in $H_{v;nn'}$
from site $j$ to site $i$.
If $n=n'$, there is no cost of Coulomb interaction to be
paid since initial and final charge configurations are the
same yet swapped, i.e., the charge configuration
changes from $(n_i, n_j) = (n-1,n)$ to $(n,n-1)$.
Conversely, 
for $n > n'$, the charge imbalance $n_i - n_j > 0$ between
sites $i$ and $j$ further increases by acting with $H_{v;nn'}$.
Therefore the Coulomb energy to be paid, $(n-n') \cdot U>0$, is positive.

Due to the specific structure of the Coulomb
interaction in the Hubbard model, the prefactor on the r.h.s.~of \Eq{eq:HU:comm} only depends on the difference
$n-n'$. Hence we can further group
the terms $H_{v;nn'}$ with the same $m \equiv n-n'$,
resulting in%
\begin{subequations}
\begin{eqnarray}
H_v = 
\sum_{m = -(N-1)}^{N-1} \underbrace{\sum_{n} H_{v;n,n-m}}_{\equiv H_{v;m}}
\label{eq:Hvm} 
\end{eqnarray}
where
\begin{eqnarray}
[H_U, H_{v;m}] &=& mU \cdot H_{v;m}, \label{eqHvm:CR} \\
H_{v;m}^\dagger &=& H_{v;-m}.
\end{eqnarray}
\end{subequations}
The index range $-N<m<N$ will be implied
unless specified otherwise.

Now substituting \Eq{eqHvm:CR} into \Eq{eq:Hrot},
we obtain the rotating frame Hamiltonian
\begin{equation}
H_\mr{rot} = H_\mu + \sum_{m} H_{v;m} e^{imUt}
\text{.}\label{eq:Hrotm}
\end{equation}
This is periodic in time with the driving frequencies given by harmonics of the
Coulomb interaction $U$.

\subsubsection{Effective Hamiltonian}

The last step of the generalized SWT is to average out fast
dynamics at frequency scale $\gtrsim U$ within the rotating frame Hamiltonian $H_\mr{rot}$.
Thus one obtains an effective Hamiltonian $H_\mr{eff}$ that describes slow, i.e., non-stroboscopic dynamics.
This $H_\mr{eff}$ is derived by applying the high-frequency expansion to $H_\mr{rot}$~\cite{Bukov2016,Bukov2015}.
(In contrast, an effective Hamiltonian given by the Magnus expansion generates the dynamics at exactly stroboscopic times, e.g., multiples of $2\pi / U$~\cite{Bukov2015}.)
Then we expand the Hamiltonian as a power series in the inverse
frequency $1/U$, with the result
\begin{equation}
\begin{aligned}
H_\mr{eff} &= H_\mu + H_{v;0} + H_{v^2/U} + O ({v^3}/{U^2}), \\
H_{v^2 \!/ U} &\equiv \hspace{-.1ex} \sum_{m \neq 0} \hspace{-.1ex} \frac{H_{v;m} H_{v;-m}}{mU} 
= \hspace{-.1ex} \sum_{m>0} \hspace{-.1ex} \frac{[H_{v;m}, H_{v;-m}]}{mU} \text{ .}
\end{aligned}
\label{eq:Heff:gen} 
\end{equation}
Essentially, the newly generated term $H_{v^2/U}$ resembles
second order perturbation theory in $v/U$, yet with subtle twists
(see discussion in \Sec{sec:SWT_interpretation}).
Via the commutator structure 
in \Eq{eq:Heff:gen},
only those terms in $H_{v^2 / U}$ survive
the lattice sums in $H_{v;m}$ and $H_{v;-m}$,
for which the pair-wise nearest-neighbor hoppings
overlap with respect to the sites they act upon.
Hereafter we will neglect the term of order $O(v^3 / U^2)$.

The non-stroboscopic time evolution of operators $\tilde{c}_{i\nu;n}$
in the original lab frame
is equivalent to the time evolution, generated by $H_\mr{eff}$,
of the dressed operators $\overline{\tilde{c}_{i\nu;n}}$
which are obtained by averaging out the fast motion of $\tilde{c}_{i\nu;n}$~\cite{Bukov2016,Bukov2015}.
We find that $\overline{\tilde{c}_{i\nu;n}} = \tilde{c}_{i\nu;n} + O(v^2/U^2)$, i.e.,
the correction to $\tilde{c}_{i\nu;n}$ has the same order as the term in $H_\mr{eff} / v$ to be neglected.
Hence we can consistently neglect both high-order terms, i.e.,
$H_\mr{eff} \approx H_\mu + H_{v;0} + H_{v^2/U}$ and $\overline{\tilde{c}_{i\nu;n}} \approx \tilde{c}_{i\nu;n}$,
to describe the original non-stroboscopic dynamics of $\tilde{c}_{i\nu;n}$
up to order $O(v/U)$.

\subsubsection{Pair hopping and symmetry
\label{sec:Heff_symm}}

  The effective low-energy Hamiltonian $H_\mr{eff}$ cannot break
  the symmetries present in the original Hamiltonian $H$.
  Here for the $\mr{SU}(N)$ Hubbard model, these are
  $U(1)_\text{charge}$ symmetry and 
  $\mr{SU}(N)_\text{flavor}$ symmetry for general $N$.
  All parts that constitute $H_\mr{eff}$, therefore, also
  must respect these symmetries.

The elementary building block in $H_v$
  in \Eq{eq:Hvsplit},
is the dimensionless operator,
\begin{eqnarray}
      \Pi_{ij}^{nm} \equiv \tfrac{v_{ij}}{v} \sum_\nu 
      \tilde{c}_{i\nu;n}^\dagger \tilde{c}_{j\nu;n-m} ,
\label{eq:Pij}
\end{eqnarray}
which encodes the phase of the hopping amplitude $v_{ij}$
  (with $v$ taken real and positive),
  as well as the nearest-neighbor lattice structure of the 
  Hamiltonian. Due to
\begin{eqnarray}
H_{v;m} = v \sum_{i j; n} \Pi_{ij}^{nm}
\text{,} \label{eq:Hvm:Pi}
\end{eqnarray}
the operator $\Pi_{ij}^{nm}$ constitutes
the kinetic energy $H_v$ in \Eq{eq:Hvm},
and thus subsequently also the effective interaction
$H_{v^2/U}$ in \Eq{eq:Heff:gen}. 
$\Pi_{ij}^{nm}$ preserves the
$\mr{U}(1)_{\mathrm{charge}} \otimes \mr{SU}(N)_\text{flavor}$
symmetry, and therefore represents
a scalar operator w.r.t.\ these symmetries.
Nevertheless, it is a non-Hermitian operator in
that it describes the directed and projected hopping
process from site $j$ to site $i$.

For the projected hopping $\Pi_{ij}^{nm}$ in \Eq{eq:Hvm:Pi},
the case $m=1$ is special in that 
the state on which $\Pi_{ij}^{n,m=1}$ acts
must have a component with equal occupation $n-1$ on both sites, $i$ and $j$.
This operator can be symmetrized 
w.r.t.\ the lattice sites,
\begin{subequations}
  \label{eq:Pi:ij:symm}
  \begin{eqnarray}
  \Pi_{ij;S}^{n,1} &\equiv& \tfrac{1}{2}
  \bigl(\Pi_{ij}^{n,1} + \Pi_{ji}^{n,1} \bigr)
  \label{eq:Pi:ij:symm:a} \\
  &=& \tfrac{1}{2}
    \sum_\nu \bigl( c_{i \nu}^\dagger c_{j \nu}
    + c_{j \nu}^\dagger c_{i \nu} \bigr)
  \cdot P_{i,n-1} P_{j,n-1}
  \notag \\
  \Pi_{ij;S}^{\bar{n}+1,1} &\overset{(\ref{eq:dh:gen})}{=}&
  \tfrac{1}{2}
  \sum_\nu \bigl(
  d_{i \nu}^\dagger h_{j \nu}^\dagger
  + d_{j \nu}^\dagger h_{i \nu}^\dagger \bigr)
  \text{,} \label{eq:Pi:ij:symm:b}
  \end{eqnarray}
\end{subequations}
where in the last line $\bar{n}=n-1$
represents the average integer filling.
Now when acting on an initial state $|\psi\rangle$,
the operator $\Pi_{ij;S}^{\bar{n}+1,1}$ first projects
into the charge sector of $\bar{n}$ particles on both
sites $i$ and $j$, i.e.,
$|\psi'\rangle \equiv P_{i,\bar{n}} P_{j,\bar{n}} |\psi\rangle$
and then generates a nearest-neighbor particle-hole
excitation, i.e., a doublon-holon pair,
by transferring one particle from site $j$
to site $i$ or vice versa.
Note that $\Pi_{ij;S}^{n,1}$ is still non-Hermitian.

For the case $N=2$ with particle-hole symmetry,
the symmetrized operator $\Pi_{ij;S}^{2,1}$
[i.e., having $\bar{n} = 1$ in \Eq{eq:Pi:ij:symm:b}]
generates a singlet in the
particle-hole sector, 
and thus respects larger symmetry:
the $\mr{SU}(2)_\text{charge}$ symmetry.
For a single site,
half-filled states are singlets w.r.t.\ the
$\mr{SU}(2)_\text{charge}$ symmetry, 
i.e., have charge quantum number $C=0$,
whereas doublon and holon states represent a doublet
with $C = 1/2$~\cite{Weichselbaum2012:sym}.
Taking the half-filled case for two sites then,
using standard spin-notation,
i.e., $\nu\in\{ \uparrow, \downarrow \}$,
we may start with the 
spin-singlet
\begin{subequations}
\label{eq:SU2:singlets}
\begin{eqnarray}
\ket{S_{ij}} \equiv \tfrac{1}{\sqrt{2}} (
  c_{i\uparrow}^\dagger c_{j\downarrow}^\dagger - 
  c_{i\downarrow}^\dagger c_{j\uparrow}^\dagger
) |0\rangle ,
\label{eq:sp:singlet}
\end{eqnarray}
with
$|0\rangle $ the vacuum state with no particles.
From the above, $| S_{ij} \rangle$
  is also a charge-singlet. Then the creation
  of a particle-hole pair, i.e.,
\begin{equation}
| C_{ij} \rangle  \equiv
\Pi_{ij;S}^{n,1} | S_{ij} 
\rangle =  \tfrac{1}{\sqrt{2}}
( c_{i\uparrow}^\dagger c_{i\downarrow}^\dagger - 
  c_{j\downarrow}^\dagger c_{j\uparrow}^\dagger
) |0\rangle ,
\label{eq:ph:singlet}
\end{equation}
\end{subequations}
still yields a singlet, both in the $\mr{SU}(2)_\text{charge}$ as well as $\mr{SU}(2)_\text{spin}$ symmetry
[note that the raising operator
$C_{i}^+ \equiv s_i c_{i\uparrow}^\dagger c_{i\downarrow}^\dagger$
within $\mr{SU}(2)_\text{charge}$
comes with alternating sign factors $s_i$
on a bipartite lattice,
such that nearest neighbours have opposite sign~\cite{Weichselbaum2012:sym};
then with
$C_{\mathrm{tot}}^+ \equiv C_{i}^+ + C_{j}^+$
and the lowering operators
$C_{i}^- \equiv (C_{i}^+)^\dagger$,
given $|C_{ij} \rangle \propto (C_{i}^+ - C_{j}^+) |0\rangle $,
it holds $C_{\mathrm{tot}}^{\pm} | C_{ij} \rangle = 0$].

From this we conclude that for the specific case \mbox{$m=1$},
in contrast to $\Pi_{ij}^{n,1}$, it is
the symmetrized operator $\Pi_{ij;S}^{n,1}$ in \Eq{eq:Pi:ij:symm}
that has scalar symmetry character 
in the spin sector and for $N = 2$ also in the charge sector,
and thus respects larger symmetry. It is in this sense that
we consider the symmetrized operator $\Pi_{ij;S}^{n,1}$
in \Eq{eq:Pi:ij:symm} more suitable to define a simple scalar
order parameter also for general $N$
(see Sec.~\ref{sec:DHcorrel} below).

\subsection{Effective Interactions}
\label{sec:eff_int}

The term $H_{v^2/U}$ in \Eq{eq:Heff:gen} includes
four types of second-order processes: two 2-site 
and two 3-site processes. The 2-site processes are
(i) hopping back and forth without actual particle transfer
which leads to an $\vec{S}\cdot \vec{S}$ (ss) type flavor-flavor
interaction, and (ii)
hopping of a pair of particles between nearest-neighbor sites
which relates to doublon-holon (dh) dynamics.
The 3-site processes appear on three neighboring sites in that two of them (say $j\neq k$)
are nearest neighbors of site $i$. 
Then the 3-site processes consist of 
(iii) hopping $j \to i$ and hopping $i \to k$, resulting in 
a correlated hopping (coh) 
of a particle from site $j$ to $k$ that depends on the state
of site $i$, and
(iv) creation (annihilation) of a pair of particles at site $i$
originating from (splitting towards) sites
$j$ and $k$, respectively. The latter represents two processes
which are hermitian conjugates of each other, i.e.,
doublon-holon creation and annihilation (dhx).
Therefore, overall, we have
\begin{subequations}
\begin{align}
  H_{v^2/U} &= H_{ss} + H_{dh} +
  \underbrace{H_\mr{coh} + H_\mr{dhx}}_{\equiv H_\text{3-site}}
  \text{.}\label{eq:Hvu}
\end{align}
By now, for simplicity, we also can role back the
commutator in \Eq{eq:Heff:gen} to a plain product by
reintroducing $m<0$ in the sum, which leads to
\begin{align}
H_{ss} &= \hspace{-.5ex}
      \sum_{ \substack{i,j;n \\ m \neq 0,1}}
      \tfrac{v^2}{mU} \cdot
      \Pi_{ij}^{nm} (\Pi_{ij}^{nm})^\dagger
\label{eq:Hss} \\
   &\equiv \hspace{-.5ex} \sum_{ \substack{i,j;n \\ m \neq 0,1}}
      \tfrac{v^2}{mU} \left(-2 \vec{\mc{S}}_i \cdot \vec{\mc{S}}_j + 
      \tfrac{\hat{n}_i (N - \hat{n}_j)}{N} \right) P_{i,n} P_{j,n-m-1}
   \nonumber \\
H_{dh} &=
  \tfrac{v^2}{2U} \sum_{i,j;n} 
  (\Pi_{ij}^{n,1} + \Pi_{ji}^{n,1}) \cdot (\Pi_{ij}^{n,1} + \Pi_{ji}^{n,1})^\dagger
\nonumber \\
  &\!\!\! \overset{\eqref{eq:Pi:ij:symm}}{\equiv} \!
   \tfrac{2 v^2}{U} \sum_{i,j;n} \Pi_{ij;S}^{n,1}
  (\Pi_{ij;S}^{n,1})^\dagger 
\label{eq:Hdh} \\
  H_\mr{coh} &\equiv \sum_{\substack{ i j k;n \\ m \neq 0}} \tfrac{v^2}{mU}
  \Pi_{ij}^{nm} \bigl(\Pi_{ik}^{nm} + \Pi_{kj}^{nm} \bigr)^\dagger
\label{eq:Hcoh} \\
  H_\mr{dhx} &\equiv \sum_{\substack{ i j k;n \\ m \neq 0}} \tfrac{v^2}{mU}
  \Pi_{ij}^{nm} \bigl(\Pi_{ki}^{n+m-1,m} \!+ \Pi_{jk}^{n-m+1,m} \bigr)^\dagger 
\label{eq:Hdhx}
\!\!\text{.}
\end{align}
\label{eq:Hvu_all}
\end{subequations}
Each term above will be derived and discussed next.

The effective spin-spin interaction $H_{ss}$ originates from
\begin{eqnarray}
H_{ss} &\equiv& \sum_{\substack{ ij \\ \nu\nu'}}
\sum_{\substack{ nn' \\ m \neq 0,1 }}
\tfrac{|v_{ij}|^2}{mU} \cdot
\tilde{c}_{i\nu;n}^\dagger \underbrace{ \tilde{c}_{j\nu;n-m} \cdot
  \tilde{c}_{j\nu';n'-m}^\dagger}_{\propto \delta_{nn'}} \tilde{c}_{i\nu';n'}
\notag
\text{,}
\end{eqnarray}
which together with \Eq{eq:Pij} yields \Eq{eq:Hss}.
Here the $m=1$ term has been deliberately excluded,
which may appear artificial, at first glance. After
all, it
represents a second order hopping process that leaves
the charge configuration intact and hence one may
assign to $H_{ss}$. Nevertheless, for symmetry reasons,
it will rather be associated with $H_{dh}$, as
explained below.
The flavor-flavor interaction in the first line
of \Eq{eq:Hss} can be written in terms of $\mr{SU}(N)$ spin
operators, since
\begin{align}
  & \Pi_{ij}^{nm} (\Pi_{ij}^{nm})^\dagger = \sum_{\nu\nu'} \hspace{-2ex}
  \underbrace{c_{i\nu}^\dagger c_{j\nu} c_{j\nu'}^\dagger  c_{i\nu'}}_{
    =c_{i\nu}^\dagger c_{i\nu'}
    \left(\delta_{\nu\nu'} - c_{j\nu'}^\dagger c_{j\nu} \right)}
  \hspace{-3ex}
  \cdot P_{i,n} P_{j,n-m-1} \notag \\
  &= \left(  -  2 \vec{\mc{S}}_i \cdot \vec{\mc{S}}_j + \hat{n}_i - \tfrac{1}{N} \hat{n}_i \hat{n}_j
  \right)
  \cdot P_{i,n} P_{j,n-m-1},    
  \label{eq:Hss:SdotS}
\end{align}
where $\vec{\mc{S}}_i \equiv \tfrac{1}{2}
\sum_{\nu \nu'} c_{i\nu}^\dagger [ \vec{G} ]_{\nu\nu'} c_{i\nu'}$
is the $\mr{SU}(N)$ generalization of the spin operator, and
$\vec{G} = (G_1, \ldots, G_{N^2 -1})$ is the set of $\mr{SU}(N)$ symmetry generators in the defining representation
with the conventional normalization $\tr ( G_a G_b ) = 2 \delta_{ab}$.
Here we have used the identity
\begin{equation*}
  \sum_{\nu \nu'} c_{i\nu}^\dagger c_{i\nu'} \cdot c_{j\nu'}^\dagger c_{j\nu}
  = 2 \vec{\mc{S}}_i \cdot \vec{\mc{S}}_j + \tfrac{1}{N} \hat{n}_i \hat{n}_j
  \text{,}
\end{equation*}
having $N$ symmetric flavors. This leads to the second line
in \Eq{eq:Hss}.

The effective doublon-holon term $H_{dh}$ in \Eq{eq:Hdh} originates from the pair-hopping,
\begin{eqnarray}
    \tilde{H}_{dh} &\equiv& \sum_{\substack{i j \\ \nu \nu'}}
       \sum_{\substack{n n' \\ m\neq 0}}
    \tfrac{v_{ij}^2}{mU} \cdot
    \underbrace{ \tilde{c}_{i\nu;n}^\dagger  \tilde{c}_{j\nu;n-m} \cdot
    \tilde{c}_{i\nu';n'-m}^\dagger  \tilde{c}_{j\nu';n'}}_{\propto\, \delta_{nn'}\delta_{m,1}}
    \notag \\
   & \overset{(\ref{eq:Pij})}{=}&
      \tfrac{v^2}{U} \sum_{i,j;n}
      \Pi_{ij}^{n,1} (\Pi_{ji}^{n,1})^\dagger   
\text{,}\label{eq:Hdh:0} 
\end{eqnarray}
where the tilde on $\tilde{H}_\mr{dh}$ 
indicates the initially strict constraint to the transfer
of a pair of particles.
In order for the Coulomb interaction energy before and after the second-order
process of a pair-hoping to be the same, the charge
configurations must be the same, yet reversed.
This is the underlying natural reason for obtaining the constraint $m=1$.

Now as discussed with
Eqs.~\eqref{eq:Pi:ij:symm} and \eqref{eq:SU2:singlets} above,
for this specific case of $m=1$, the individual terms
in \Eq{eq:Hdh:0} do not respect the $\mr{SU}(2)_\text{charge}$
symmetry if present.
In the same manner, as a $\vec{C}_i \cdot
\vec{C}_j$ interaction, with $\vec{C}$
the pseudo-spin operator in the particle-hole channel
for $N=2$, includes a $C_{i}^z C_{j}^z$ term that leaves
the local charge configuration intact, it is desirable
to symmetrize \Eq{eq:Hdh:0}. Therefore based on the
earlier discussion with Eqs.~\eqref{eq:Pi:ij:symm}
  and \eqref{eq:SU2:singlets},
above, we define
\begin{eqnarray}
    H_{dh} &\equiv& 
      \tilde{H}_{dh}  +
      \tfrac{v^2}{U} \sum_{i,j;n} 
      \Pi_{ij}^{n,1} (\Pi_{ij}^{n,1})^\dagger  \notag 
     = \tfrac{2 v^2}{U} \sum_{i,j;n} 
          \Pi_{ij;S}^{n,1} (\Pi_{ij;S}^{n,1})^\dagger
      \label{eq:Hdh:2} 
\text{.}
\end{eqnarray}
Here the term added to $\tilde{H}_{dh}$ exactly recovers
the $m=1$ term that has been already intentionally excluded
from \Eq{eq:Hss} above exactly for this reason.

For the case $N=2$ with particle-hole symmetry then,
$H_{ss}$ and $H_{dh}$ reduce to~\cite{Lee2017}
\begin{align}
& H_{ss} = \frac{v^2}{U} \sum_{\la i,j \ra}
4 \vec{S}_i \cdot \vec{S}_j - P_{i1} P_{j1}, \nonumber \\
& H_{dh} = \frac{2 v^2}{U} \sum_{\la i,j \ra} 
 (c_{j1}^\dagger c_{j2}^\dagger c_{i2} c_{i1} + P_{i2} P_{j0}) + (i \leftrightarrow j) \nonumber \\
 & \phantom{H_{02}} = \frac{v^2}{U} \sum_{\la i,j \ra, \nu, \nu'} 
( h_{i\nu}^\dagger d_{j\nu}^\dagger + h_{j\nu}^\dagger d_{i\nu}^\dagger ) 
( d_{i\nu'} h_{j\nu'} + d_{j\nu'} h_{i\nu'} ) , \nonumber 
\end{align}%
where the flavor operator $\vec{\mc{S}}_i$ reduces to
the standard $\mr{SU}(2)$ spin operator $\vec{S}_i$,
and $\la i,j \ra$ indicates nearest neighbour pairs of sites.
In this case,
the only remaining terms are $n=1$ and $m=-1$ 
for \Eq{eq:Hss}, and $n = \bar{n}+1 = 2$ for
\Eq{eq:Hdh}.
Here $H_{ss}$ and $H_{dh}$ can also be written as the
  projectors
\begin{subequations}
\begin{align}
 H_{ss} &= \tfrac{-4v^2}{U} \sum_{\la i,j \ra} \prj{S_{ij}}, \\
 H_{dh} &= \tfrac{4v^2}{U} \sum_{\la i,j \ra} \prj{C_{ij}},
\end{align}%
\label{eq:Hss_Hdh_SU2}%
\end{subequations}%
with $\ket{S_{ij}}$ and $\ket{C_{ij}}$ as defined 
  in \Eq{eq:SU2:singlets}.

The 3-site effective interactions in \Eq{eq:Hcoh} and
  \Eq{eq:Hdhx}, finally, are derived in complete analogy
  to the above without any further ado.

\subsection{Interpretation in terms of second order perturbation theory}
\label{sec:SWT_interpretation}

Though the effective Hamiltonian in \Eq{eq:Heff:gen}
is derived by employing the high-frequency expansion from Floquet theory,
it can be understood in an easier way.
Indeed, the term $H_{v;m} H_{v;-m} / mU$ 
simply describes a second-order virtual process
in which an intermediate state differs in energy from the initial and the final states by $-mU$ due 
to the Coulomb interaction $H_U$.
The way in which such second-order terms are included into the effective Hamiltonian is similar to 
what is done in the perturbation theory approach to the SWT~\cite{Kato1949,*Takahashi1977,Hewson1993}.

However, there is a subtle difference.
In the perturbation theory approach,
the intermediate state has clearly higher energy
than those of the initial and the final states.
Here, in contrast, the terms $H_{v;m} H_{v;-m} / mU$ with $m > 0$ are also incorporated 
in \Eq{eq:Heff:gen}, that is,
the intermediate state can have ``lower'' interaction energy by $- m U < 0$.
At first glance, it seems to be contradictory to the spirit of the SWT that the virtual process, 
starting from and ending at a low-energy subspace, 
should involve an intermediate state of higher energy.

To resolve this,
we remark that the denominator $mU$ is the energy
difference measured by only the local Coulomb
interaction $H_U$, not by the full Hamiltonian
that also includes kinetic energy.
The Fermi-liquid ground state in the metallic phase 
involves local charge fluctuations, which give rise to its
  metallicity. Acting with $H_{v;-m}$ ($m > 0$) onto this ground
state will decrease the Coulomb energy $\la H_U \ra$ by $m U$, but
increase the total energy $\la H \ra$, since it is not the ground
state anymore.  Thus the intermediate state implied by the terms
$H_{v;m} H_{v;-m} / mU$ of $m > 0$ (e.g., $H_{dh}$ originating from
$m = 1$) has higher energy.  On the other hand, in the insulating
phase, the ground state is mainly spanned by the basis in which
  the lattice sites are filled with the average integer occupation
  $\bar{n}$.  The contribution of doublons and holons to the ground
  state is finite but small, as shown in
    Ref.~\cite{Yokoyama1990}.  Thus, contrary to the metallic case,
the terms $H_{v;m} H_{v;-m} / mU$ of $m > 0$ become 
much less relevant
to the low-energy subspace in the insulating phase.

Therefore the summation over all possible values of $m$,
positive as well as negative, is not contradictory at all.
It is rather an unbiased way of including all second-order processes, without pre-defining any low-energy subspace.
In other words, the construction of the effective Hamiltonian
in \Eq{eq:Heff:gen} represents all possible ``slow'' processes,
i.e., dynamics within an energy window of narrow
width $\sim v^2 / U$. In contrast,
the $t$-$J$ model does not contain the doublon-holon term $H_{dh}$, since
the terms of $m > 0$ in \Eq{eq:Heff:gen} are neglected by
the assumption that the half-filled subspace is low-energy subspace.
In this sense, the $t$-$J$ model provides an 
incomplete description for the metallic phase.

\section{Mean-field decoupling scheme for
doublon and holon correlators}
\label{sec:DHcorrel}

We utilize the equations of motion approach
to compute the correlators of the projected particle operators $\tilde{c}_{i\nu;n}$
whose time evolutions are generated by the effective Hamiltonian of the
$\mr{SU}(N)$ Hubbard model.
The effective Hamiltonian is given by $H_\mr{eff}$ in
Eqs.~\eqref{eq:Heff:gen} and \eqref{eq:Hvu_all}
while neglecting terms of order $O(v^3 / U^2)$. 
The equation of motion for the correlator in frequency domain is given by
\begin{equation}
\begin{aligned}
&\omega^+ 
\la \tilde{c}_{i\nu;n} \Vert \tilde{c}_{i'\nu;n'}^\dagger \ra_\omega = \\
& \qquad \la \{ \tilde{c}_{i\nu;n}, \tilde{c}_{i'\nu;n'}^\dagger \} \ra - \la [H_\mr{eff}, \tilde{c}_{i\nu;n}] \Vert \tilde{c}_{i'\nu;n'}^\dagger \ra_\omega ,
\end{aligned}
\label{eq:EOM_omega}
\end{equation}
where $\langle A \Vert B \rangle_\omega \equiv 
\int dt \, e^{i\omega^+ t} \langle A \Vert B \rangle_t$
with \mbox{$\omega^+ \equiv \omega+i0^+$} and $\langle A \Vert B \rangle_t = G_{AB}(t) \equiv
-i\vartheta(t) \langle \{ A(t), B(0) \} \rangle_T$,
assuming fermionic operators $A$ and $B$, i.e., containing an
odd number of (projected) fermionic creation and annihilation
operators. 
For bosonic operators $A$ and $B$, such as for the charge susceptibility 
$\chi_c(\omega) \equiv \langle \delta \hat{n}_i \Vert \delta \hat{n}_i \rangle_\omega$
($\delta \hat{n}_i \equiv \hat{n}_i - \la \hat{n}_i \ra$)
the anticommutator would be replaced by the commutator $[A(t),B]$.
In \Eq{eq:EOM_omega}
we consider general locations $i$ and $i'$
in the correlator which will be required for the Fourier
transform to momentum space later.
Since $H_\mr{eff}$ is not quadratic, the 
equations of motion generated by \Eq{eq:EOM_omega}
do not close.

The mean-field and the decoupling approximations to 
close the equations of motion are as follows.
First, we regard the paramagnetic metallic ground state as the condensate
of the pair of doublon- and holon-like excitations,
e.g., created by the action of $\Pi_{ij;S}^{n1}$ [cf.~\Eq{eq:Pi:ij:symm}]
which represents a projected part
of the hopping Hamiltonian $H_v$ [see Eqs.~\eqref{eq:Hvm} and
\eqref{eq:Hvm:Pi}].
Following the discussion in Secs.~\ref{sec:Heff_symm} and \ref{sec:eff_int},
the condensate also respects charge and spin symmetries,
i.e., does not break them in the metallic phase.
(In the insulating phase, on the other hand,
the doublon-holon pair do not condense, since charge fluctuations are
largely suppressed;
see \Sec{sec:SWT_interpretation} above.)

Accordingly we define the mean-field variables
\begin{equation}
   \Delta_n \equiv
   -v \la (\Pi_{ij;S}^{n1})^\dagger \ra =
   \frac{-v}{2} \la (\Pi_{ij}^{n1} + \Pi_{ji}^{n1})^\dagger \ra
\text{,}
\end{equation}
which is positive when the expectation value is taken
with the Fermi-liquid ground state of the metallic phase.
For $U = 0$,
it is clear that the expectation value of hopping term is negative,
$\langle H_v \rangle = \sum_{ij} v_{ij} 
\langle c_i^\dagger c_j \rangle < 0$, by filling
states with negative one-particle energies.
The operator $v (\Pi_{ij;S}^{n1})^\dagger$,
which annihilates a doublon-holon pair, is nothing
but a summand of a projected hopping $H_{v;-1}$
[cf.~Eqs.~\eqref{eq:Pij} and \eqref{eq:Hvm:Pi}],
thus the expectation value $v \la (\Pi_{ij;S}^{n1})^\dagger \ra$
is also negative.
The overall sign is not changed even
in the competition between the kinetic energy $\la H_v \ra$
and the Coulomb energy $\la H_U \ra$.
Using the mean-field variable $\Delta_n$, we
rewrite \Eq{eq:Hdh},
\begin{equation}
H_{dh} \approx \frac{-v}{U} \sum_{\la i j \ra; n}
\Delta_n (\Pi_{ij}^{n1} + \Pi_{ji}^{n1}) + \text{h.c.}
\label{eq:Hdh:mf}
\end{equation}
For the $\mr{SU}(2)$ case,
there is only one type of pair excitations, namely for $n = 2$
(due to the range of $n$), i.e., $\Delta_{dh} = \Delta_{n=2}$.

Second, we decouple the flavor and charge operators from the correlators of interest,
since we focus on the subpeaks on the intermediate energy scale away from those of the quasiparticle peak and the Hubbard bands.
Charge fluctuations explore high energy scales on the order of $U$,
i.e., the region of the Hubbard bands,
which are integrated out by the generalized SWT.
Flavor fluctuations [equivalent to spin in the case of $N = 2$ flavors],
on the other hand, typically remain in the low-energy sector.
Consistently, in the local density of states
(spectral function) the spin-fluctuations predominantly
contribute to the quasiparticle peak around $\omega=0$,
whereas the Hubbard side bands are largely integrated
out except for their inner subpeak structure.
This separation of the energy scales for charge and spin fluctuations
appears as the separation of the peak positions of charge and spin susceptibilities;
see Fig.~1 of \Ref{Lee2017}.

Now the commutator $[H_\mr{eff}, \tilde{c}_{i\nu;n}]$
in \Eq{eq:EOM_omega} results in a sum of 
two- and three-site terms, say $O_i O_j$ or $O_i O_j O_k$,
which group all local operators acting on sites
$i$, $j$, and $k$, respectively.
Moreover, these local operators, say $O_i$ on site $i$,
are comprised of one, two,
or three projected particle operators $\tilde{c}_{i\nu;n}$.
The local operators made of two projected operators
can be classified into three types:
pair creation and annihilation
$(\tilde{c}_{i\nu;n} \tilde{c}_{i\nu';n+1})^{(\dagger)}$, 
spin flip $\tilde{c}_{i\nu;n}^\dagger \tilde{c}_{i\nu';n}$ ($\nu \neq \nu'$),
and charge measurement $\tilde{c}_{i\nu;n}^\dagger \tilde{c}_{i\nu;n}
\propto P_{i,n}$.
We decouple such local operators made of two projected operators,
say $O'_i$, from the correlators, e.g.,
$\la O'_i O_j \Vert O_k \ra_\omega \approx \la O'_i \ra
\la O_j \Vert O_i \ra_\omega$.
Since the metallic ground state conserves the total number
of particles of each flavor $\sum_i n_{i\nu}$
and is not flavor-polarized, only the charge measurement
type of local operator has finite expectation value.
Hence we need to expand the $\Pi$ terms in $H_{\mathrm{eff}}$
in favor of the new shortcuts,
\begin{subequations}
\label{eq:MF:decoupling}
\begin{align} 
& A_{i\nu}^{n} \equiv \tilde{c}_{i\nu;n}^\dagger \tilde{c}_{i\nu;n} \rightarrow
  \la A_{i\nu}^{n} \ra = \tfrac{n}{N} \la P_{in} \rangle
\\
& B_{i\nu}^{n} \equiv \tilde{c}_{i\nu;n} \tilde{c}_{i\nu;n}^\dagger \rightarrow
  \la B_{i\nu}^{n} \ra = \tfrac{N-n+1}{N} \la P_{i,n-1} \ra
\\
&     \tilde{c}_{i\nu;n}^\dagger \tilde{c}_{i\nu';n'} \rightarrow
  \la \tilde{c}_{i\nu;n}^\dagger \tilde{c}_{i\nu';n'} \rangle = 0
  \text{ if } \nu\neq\nu'
\\
&     \tilde{c}_{i\nu;n-1} \tilde{c}_{i\nu;n} \rightarrow
  \la \tilde{c}_{i\nu;n-1} \tilde{c}_{i\nu;n} \rangle = 0
  \text{.}
\end{align}
\end{subequations}%
Furthermore, we neglect the 3-site terms $O_i O_j O_k$
in the commutator $[H_\mr{eff}, \tilde{c}_{i\nu;n}]$
where all $O_i$, $O_j$, and $O_k$ are made of odd number of projected particle operators.
We presume that these approximations
can be compensated by renormalizing the system parameters.

With this, the first term on the r.h.s.\ of \Eq{eq:EOM_omega} becomes
\begin{align}
   \la \{ \tilde{c}_{i\nu;n}, \tilde{c}_{i'\nu;n'}^\dagger \} \ra
  &= \delta_{ii'} \delta_{nn'} \la A_{i\nu}^{n} + B_{i\nu}^{n} \ra
\label{eq:comm_approx2}
\text{.} \notag 
\end{align}
Note that the $\tilde{c}$ operators do not obey fermionic
commutator relations, hence the result here is not just given
by canonical fermionic commutator relations.

Now we expand the commutator $[H_\mr{eff}, \tilde{c}_{i\nu;n}]$
in \Eq{eq:EOM_omega} term by term.
From \Eq{eq:Heff:gen}, we have
\begin{align}
     [ H_\mu, \tilde{c}_{i\nu;n} ] &=
     -\mu \sum_{i n'}  [ n' P_{in'}, \tilde{c}_{i\nu;n} ]
     = \mu \tilde{c}_{i\nu;n} ,
\\
     [ H_{v,0}, \tilde{c}_{i\nu;n} ] &=
     \sum_{i'j'\nu'; n'} v_{i'j'} \underbrace{
     [ \Pi_{i'j'}^{n',0},
     \tilde{c}_{i\nu;n} ]
     }_{\propto \ \delta_{i'i} \delta_{\nu'\nu} + \, \ldots } \notag \\
     &\approx - \la A_{i\nu}^{n} + B_{i\nu}^{n} \ra
     \sum_{j} v_{ij} \tilde{c}_{j\nu;n} , \label{eq:comm:Hv0}
\end{align}
where $\ldots$ means the terms to be neglected 
via the decoupling approximation in \Eq{eq:MF:decoupling}.
The sum $\la A_{i\nu}^{n} + B_{i\nu}^{n} \ra$ in \Eq{eq:comm:Hv0} is the remnant of the commutator.
\begin{widetext}
Similarly for the doublon-holon term,
\begin{align}
   & [ H_{dh}, \tilde{c}_{i\nu;n} ]   \approx \frac{-v}{U} \sum_{j n'}
   \bigl[ \Delta_{n'} \Pi_{ij}^{n'1} + \Delta_{n'}^\ast (\Pi_{ji}^{n'1})^\dagger
        + \ldots, \tilde{c}_{i\nu;n} 
   \bigr] 
	\approx
	\la A_{i\nu}^{n} + B_{i\nu}^{n} \ra 
	  \sum_{j} v_{ij}
	   \Bigr(
	     \tfrac{\Delta_n}{U} \tilde{c}_{j\nu;n-1}
	   + \tfrac{\Delta_{n+1}^\ast}{U} \tilde{c}_{j\nu;n+1}
	   \Bigl)
\text{,}\label{eq:Hdh:comm}
\end{align}
and for the flavor-flavor interactions,
\begin{align}
   & [ H_{ss}, \tilde{c}_{i\nu;n} ] =
     \sum_{ j n'} \sum_{ m \neq 0,1}
      \tfrac{v^2}{mU} 
     \Bigl[ \Pi_{ij}^{n'm} (\Pi_{ij}^{n'm})^\dagger +
            \Pi_{ji}^{n'm} (\Pi_{ji}^{n'm})^\dagger, \tilde{c}_{i\nu;n}
     \Bigr] \notag \\
   & = \sum_{ j\nu'} \sum_{ m\neq 0,1} \tfrac{|v_{ij}|^2}{mU} 
    \Bigl(
        \bigl(
            A_{i\nu'}^{n-1} B_{j\nu'}^{n-m-1} 
          + B_{i\nu'}^{n} A_{j\nu'}^{n+m}
        \bigr) \tilde{c}_{i\nu;n} -
        \tilde{c}_{i\nu;n}
        \bigl(
            A_{i\nu'}^{n} B_{j\nu'}^{n-m} 
          + B_{i\nu'}^{n+1} A_{j\nu'}^{n+m+1} 
        \bigr) \Bigr) + \ldots \notag \\
   & \overset{\eqref{eq:MF:decoupling}}{\approx}
    \sum_{j} \sum_{ m\neq 0,1} \tfrac{|v_{ij}|^2}{mU} 
    \Bigl(
            (n-1)\la B_{j\nu^{(\prime)}}^{n-m-1} \ra
          + (N-n+1) \la A_{j\nu^{(\prime)}}^{n+m} \ra 
          - n \la B_{j\nu^{(\prime)}}^{n-m} \ra
          - (N-n) \la A_{j\nu^{(\prime)}}^{n+m+1} \ra 
        \Bigr)
        \tilde{c}_{i\nu;n}
\text{,}\label{eq:Hss:comm}
\end{align}
where the sum over $\nu'$ survives the commutator in the first line, since the $\Pi$'s include projectors.
The mean field averaging in the third line is performed
on the quadratic operators $A$ and $B$ at site $j$ only.
Consequently, with
$\la B_{j\nu^{(\prime)}}^{n'} \ra$ assumed independent of $\nu'$
(hence the brackets around the prime),
the sum over $\nu'$ can be performed for the three operators
acting on site $i$: e.g., $\sum_{\nu'} A_{i\nu'}^{n-1}
\tilde{c}_{i\nu;n} = \sum_{\nu'} c_{i\nu'}^\dagger c_{i\nu'}
\cdot P_{i,n-1} c_{i\nu} = (n-1) \tilde{c}_{i\nu;n}$.

Finally for the 3-site term, with \Eq{eq:Hvu}, we have
\begin{align}
   & [ H_{\text{3-site}}, \tilde{c}_{i\nu;n} ]
   = [ H_\mr{coh} + \ldots, 
     \tilde{c}_{i\nu;n} ]
   \overset{\eqref{eq:Hcoh}}{=}
     \sum_{k, j \neq i; n'} \sum_{m \neq 0} \tfrac{v^2}{mU}
     \Bigl[
        \Pi_{k j}^{n'm} (\Pi_{k i}^{n'm})^\dagger
      + \Pi_{i k}^{n'm} (\Pi_{j k}^{n'm} )^\dagger
      + \ldots, \tilde{c}_{i\nu;n}
     \Bigr] \notag \\
   & =
     \sum_{k, j \neq i} \sum_{m \neq 0} \tfrac{v_{ik} v_{kj}}{mU}
    \Bigl(
        \tilde{c}_{j\nu;n} A_{k\nu}^{n+m} (A_{i\nu}^{n} + B_{i\nu}^{n}) 
      - \tilde{c}_{j\nu;n} B_{k\nu}^{n-m} (A_{i\nu}^{n} + B_{i\nu}^{n}) 
     \Bigr) + \ldots \notag \\
   & \approx
     \sum_{k, j \neq i} \sum_{m \neq 0} \tfrac{v_{ik} v_{kj}}{mU}
     \, \tilde{c}_{j\nu;n} 
         \la A_{k\nu}^{n+m} + B_{k\nu}^{n+m} \ra 
         \la A_{i\nu}^{n} + B_{i\nu}^{n} \ra  ,
\label{eq:H3site:comm}
\end{align}
where for the last line we relabeled $m\to-m$ on the
second part of the second line.
With the approximations above,
the system of equations of motion is in closed form now.

\subsection{The case of $N = 2$}

In this subsection, 
we consider the example of the $\mr{SU}(2)$ Hubbard model at half filling.
It holds by symmetry then that $\mu=0$, and
 $\la P_{i0} \ra = \la P_{i2} \ra = (1- \la P_{i1} \ra)/2$.
Therefore $\la A_{i\nu}^{n} + B_{i\nu}^{n} \ra = \tfrac{1}{2}$
for $n=1,2$.
The projected particle operators are
doublon and holon operators,
$d_{i\nu} = \tilde{c}_{i\nu;n=2}$, $h_{i\nu}^\dagger = \tilde{c}_{i\nu;n=1}$,
and there is only one mean-field variable $\Delta_{n=2} = \Delta_{dh}$.
From the above derivation for general $N$, we obtain
\begin{subequations}
\begin{align}
-[H_\mr{eff}, d_{i\nu}]
  &\approx  \sum_{j} \frac{v_{ij}}{2} \Big[
     d_{j\nu} -  \frac{\Delta_{dh}}{U} h_{j\nu}^\dagger
   + \frac{ 2 v_{ji}}{U} \la P_{i1} \ra d_{i\nu}
  \Big]
+ \sum_{k, j \neq i}
  \frac{v_{ik} v_{kj}}{4U} d_{j\nu} ,  \\
-[H_\mr{eff}, h_{i\nu}^\dagger] 
  &\approx  \sum_{j} \frac{v_{ij}}{2} \Big[
     h_{j\nu}^\dagger - \frac{{\Delta}_{dh}^*}{U} d_{j\nu} 
   - \frac{ 2 v_{ji}}{U} \la P_{i1} \ra h_{i\nu}^\dagger
  \Big]
- \sum_{k, j \neq i} \frac{v_{ik} v_{kj}}{4U} h_{j\nu}^\dagger
\text{.} 
\end{align}%
\label{eq:comm_approx_SU2}%
\end{subequations}%
\end{widetext}%
In this system of equations,
only the doublon and holon operators of the same flavor index $\nu$ are related;
hereafter, we drop the index $\nu$ for simplicity.

For the DMFT calculations in \Ref{Lee2017},
we considered a lattice which has semi-elliptic density of states, $\rho_0 (\omega) = \tfrac{2}{\pi D^2} \sqrt{D^2 - \omega^2}$.
The Bethe lattice
has such a density of states, but solving the system of equations in Eq.~\eqref{eq:comm_approx_SU2} on the Bethe lattice is not simple;
the Fourier transform is not applicable to the Bethe lattice
since it is not translationally invariant in Euclidean space.
Instead, we consider the hypercubic lattice in infinite dimensions (where one can use the Fourier transform) which approximates the semi-elliptic density of states for small energies.
The density of states for such hypercubic lattice is Gaussian~\cite{Metzner1989},
\begin{equation}
\rho_\mr{hc} ( \omega )
= \frac{1}{v \sqrt{2 \pi z }} e^{- (\omega / v \sqrt{2 z})^2},
\end{equation}
where $v$ scales as $1/\sqrt{z}$ depending on coordination number $z \to \infty$.
Here we set $D \equiv v \sqrt{z}$
(contrary to $D = 2v\sqrt{z}$ for the semi-elliptic $\rho_0 (\omega)$)
so that two density of states $\rho_0 (\omega)$ and $\rho_\mr{hc} (\omega)$ are the same for small energies $|\omega | \ll D$ up to overall factor,
\begin{equation}
\begin{aligned}
\rho_0 (\omega) &= \frac{2}{\pi D} \left( 1 - \frac{\omega^2}{2 D^2} \right) + O (\omega^4 / D^4), \\
\rho_\mr{hc} (\omega) &= \frac{1}{\sqrt{2 \pi} D} \left( 1 - \frac{\omega^2}{2 D^2} \right) + O (\omega^4 / D^4).
\end{aligned}
\end{equation}

We introduce the doublon and holon operators in momentum space,
\begin{equation}
d_{\mb{p}\alpha} \equiv \sqrt{\tfrac{2}{N_s}} \sum_{i\in \alpha} e^{-i \mb{p} \cdot \mb{r}_i} d_{i}, \,\,\,
h_{\mb{p}\alpha} \equiv \sqrt{\tfrac{2}{N_s}} \sum_{i\in \alpha} e^{-i \mb{p} \cdot \mb{r}_i} h_{i},
\nonumber
\end{equation}
where we have dropped the flavor index $\nu$
as noted after \Eq{eq:comm_approx_SU2}. 
Here $\mb{p}$ is the linear momentum
(using units of $\hbar=1$), 
$N_s$ the total number of lattice sites,
and $\mb{r}_i$ the location of site $i$.
Given the nearest neighbor hopping on a hyper-cubic
lattice, the hopping connects two bipartite sublattices,
labeled $\alpha = \mr{A}, \mr{B}$.

\begin{figure}
\centerline{\includegraphics[width=.49\textwidth]{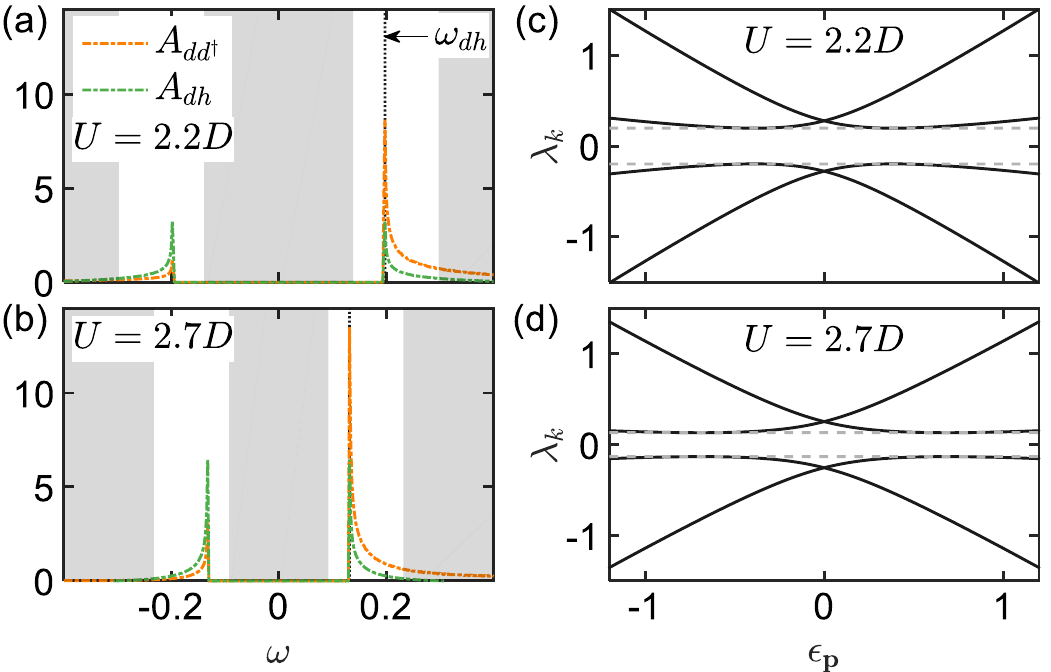}}
\caption{
Mean-field analysis of the SU($N=2$) Hubbard model.
(a), (b) The correlation functions of doublons and holons, $A_{dd^\dagger} (\omega)$ and $A_{dh} (\omega)$, have peaks at $\omega = \pm \omega_{dh}$.
Grey shading at higher (lower) energy windows
indicate that large (small) energy
scales have been neglected
by the specific type of mean-field and decoupling approximations employed on top of the generalized SWT.
(c), (d) The eigenvalues $\lambda_k$ of $H_\mb{p}$ in \Eq{eq:EOM_Hp}.
Dashed vertical lines indicate the locations of $\lambda_k = \pm \omega_{dh}$.
Each row of panels are for the same value of $U$.
Here we used a 
fixed value of $\Delta_{dh} = 2.91D$, which is equal to the critical interaction strength $U_{c2}$,
and the $U$-dependent values of $\la P_{i1} \ra$ taken from
the DMFT calculations in \Ref{Lee2017}.
}
\label{fig:eff2}
\end{figure}

We now rephrase \Eq{eq:comm_approx_SU2} in matrix form.
With 
\begin{subequations}
\begin{align}
  & \vec{O}_\mb{p} \equiv (
        d_{\mb{p}\mr{A}},
        h_{-\mb{p}\mr{A}}^\dagger,
        d_{\mb{p}\mr{B}},
        h_{-\mb{p}\mr{B}}^\dagger
    )
\\
    & 
    [G_\mb{p} (\omega)]_{mn} \equiv
    \la O_{\mb{p}m} \Vert O_{\mb{p}n}^\dagger \ra_\omega,
\end{align}
\label{eq:Gindex}%
\end{subequations}%
and using the relations for a periodic inversion symmetric lattice
\begin{subequations}
\begin{align}
& \epsilon_\mb{p} \equiv
\sum_{j} v_{ij} e^{i\mb{p}\cdot(\mb{r}_j - \mb{r}_i)} = \epsilon_{-\mb{p}}
\\
& \epsilon_\mb{p}^2 - D^2 = 
\sum_{j,k \neq i} v_{ij} v_{jk} e^{i\mb{p}\cdot(\mb{r}_k - \mb{r}_i)}
\text{,}
\end{align}
\end{subequations}
we obtain (where scalar numbers are implicitly multiplied
by the 4-dimensional identity matrix $I_{4\times 4}$)
\begin{subequations}
\begin{equation}
   [ \, \omega^+ - H_\mb{p} \, ] \, G_{\mb{p}} (\omega) = \tfrac{1}{2},
   \label{eq:GFmatrix}
\end{equation}
where 
\begin{align}
H_\mb{p} &\equiv
\frac{\epsilon_\mb{p}}{2}
\begin{pmatrix}
 & & 1 & \\
& & & 1 \\
1 & &  & \\
& 1 & & 
\end{pmatrix}
- \frac{\epsilon_\mb{p}}{2U}
\begin{pmatrix}
 & &  & {\Delta}^*_{dh} \\
& & \Delta_{dh} &  \\
 & {\Delta}^*_{dh} &  & \\
\Delta_{dh} &  & & 
\end{pmatrix} \notag \\
& +
\frac{D^2 (4 \la P_{i1} \ra - 1) + \epsilon_\mb{p}^2}{4U}
\begin{pmatrix}
1 & & & \\
& -1 & & \\
& & 1 & \\
& & & -1
\end{pmatrix}
\text{.}
\label{eq:EOM_Hp}
\end{align}%
\label{eq:EOM_matrix}%
\end{subequations}
with zero-matrix elements not shown, for readability.
From this matrix equation, we compute the local correlators as follows.
For an arbitrary but fixed value of non-interacting kinetic energy $\epsilon_\mb{p}$,
we diagonalize $H_\mb{p}$ to obtain energy eigenvalues $\lambda_k$ and corresponding eigenvectors $\vec{u}_k$ ($k = 1,\ldots, 4$).
\FIG{fig:eff2} shows for 
how $\lambda_k$ depends on $U$ and $\epsilon_\mb{p}$.
Then, for each $k$, the $(m,n)$ element of the
$4\times 4$ matrix $\vec{u}_k \vec{u}_k^T / 2$ gives
the contribution to the imaginary part of
the correlators, i.e., $[A_\mb{p}]_{mn} \equiv \tfrac{-1}{\pi} \im [G_\mb{p}]_{mn}$, at energy $\omega = \lambda_k$
as follows:
\begin{equation}
[A_\mb{p} (\omega)]_{mn} = \sum_{k = 1}^4 
\left[ \frac{\vec{u}_k (\epsilon_\mb{p}) \vec{u}_k^T (\epsilon_\mb{p})}{2} \right]_{mn} \delta ( \omega - \lambda_k (\epsilon_\mb{p})),
\nonumber
\end{equation}
where the factor $1/2$ to the matrix $\vec{u}_k \vec{u}_k^T$ originates from the r.h.s.~of \Eq{eq:GFmatrix},
and the indices $m$ and $n$ are defined by \Eq{eq:Gindex}.
With this,
we obtain the local correlator in real space, say at site $i$,
\begin{eqnarray}
&& [A_i (\omega)]_{mn} \equiv \tfrac{-1}{\pi} \im \la O_{im} \Vert O_{in}^\dagger \ra_\omega 
\label{eq:meanfield} \\
&&= \int_{-\infty}^{\infty} \mr{d} \mb{p} \, [A_\mb{p} (\omega)]_{mn}
= \int_{-\infty}^{\infty} \mr{d} \epsilon_\mb{p} \, \rho_\mr{hc} (\epsilon_\mb{p}) \, [A_\mb{p} (\omega)]_{mn}
\nonumber \\
&&= \int_{-\infty}^{\infty} \mr{d} \epsilon_\mb{p}  \sum_{k = 1}^4 
\left[ \frac{\vec{u}_k (\epsilon_\mb{p}) \vec{u}_k^T (\epsilon_\mb{p})}{2} \right]_{mn} \rho_\mr{hc} (\epsilon_\mb{p}) \, \delta ( \omega - \lambda_k (\epsilon_\mb{p}) )
\nonumber \text{,}
\end{eqnarray}
where $\vec{O}_i$ is the same as in \Eq{eq:Gindex},
except the replacement $\pm \mb{p} \to i$.
In practice, we solve the analytical expressions
in \Eq{eq:meanfield} numerically. For this,
(i) we take a grid for $\epsilon_\mb{p}$, and replace the integral with
a numerical summation.
(ii) We diagonalize the matrix $H_\mb{p}$ for each value of $\epsilon_\mb{p}$.
Then the contribution to $[A_\mb{p} (\omega)]_{mn}$ will be given as the collection of $\delta$ functions at $\lambda_k(\epsilon_\mb{p})$.
Finally, (iii) we replace the $\delta$ functions in frequency space with the Gaussians of a finite but narrow width that just interpolates the discrete intervals.

The components $[A_\mb{p} (\omega)]_{mn}$ with
$(m,n) = (1,1)$ and $(2,1)$ (or equivalently $(m,n) = (3,3)$ and $(4,3)$)
are equivalent to the spectral functions
$A_{dd^\dagger} (\omega) \equiv -\tfrac{1}{\pi}
\mathrm{Im} \la d_{i\nu} \Vert d_{i\nu}^\dagger \ra_\omega$ and
$A_{h^\dagger d^\dagger} (\omega) \equiv
-\tfrac{1}{\pi} \mathrm{Im} \la
h_{i\nu}^\dagger \Vert d_{i\nu}^\dagger \ra_\omega$,
respectively, as shown in \Fig{fig:eff2}.
They exhibit peaks at $\omega = \pm \omega_{dh}$.
Due to the operator identity 
$c_{i\nu} = d_{i\nu} + h^\dagger_{i\nu}$
for the case $N=2$,
the sum of the correlation functions of doublons and holons is equivalent to the Green's function of the particle,
that is,
\begin{equation}
A (\omega) = A_{dd^\dagger} (\omega) + A_{h^\dagger d^\dagger} (\omega) + A_{dh} (\omega) + A_{h^\dagger h} (\omega) ,
\label{eq:Adecomposition}
\end{equation}
where $A(\omega)$ is the local spectral function.
In given $\mr{SU}(2)$ case,
the particle-hole symmetry results in the symmetry of the correlation functions,
$A_{dd^\dagger} (\omega) = A_{h^\dagger h} (-\omega)$ and $A_{h^\dagger d^\dagger} (\omega) = A_{h^\dagger d^\dagger} (-\omega) = A_{dh} (\omega) = A_{dh} (-\omega)$.
Thus the peaks of correlation function $A_{dd^\dagger} (\omega)$ and $A_{h^\dagger d^\dagger} (\omega)$ directly correspond to the peak features in $A (\omega)$.

The mean-field decoupling scheme 
underlying \Eq{eq:EOM_matrix}
above deserves a few comments. First, by the 
specific approximations taken, we cut out both
the low-energy spin dynamics and high-energy charge dynamics. 
Therefore, in contrast to the numerical results presented in \Ref{Lee2017},
the analytically obtained curves in \Fig{fig:eff2}
capture neither the low-energy 
peak at the Fermi energy near $\omega = 0$
related to spin-dynamics, nor the Hubbard bands,
as indicated by the grey shaded areas in \Fig{fig:eff2}.

Second, the mean-field decoupling scheme as
introduced above is at the level of equations of motion
for correlations functions that take {\it two} parameters,
namely $\la P_{i1} \ra$ and $\Delta_{dh}$, as input.
Solving these mean-field equations for the correlation
functions does not offer a simple way to recompute
$\la P_{i1} \ra$ and $\Delta_{dh}$,
e.g., to determine these parameters self-consistently,
contrary to other mean-field approaches such as the Bardeen-Cooper-Schrieffer (BCS) theory of superconductivity.
Instead, the values of $\la P_{i1} \ra$ and $\Delta_{dh}$
\emph{do} depend on the dynamics at the decoupled energy scales.
For example, our first parameter $\la P_{i1} \ra $ is directly
related to the charge susceptibility
$\chi_c (\omega)$ at all frequencies
since, e.g., at half filling and $T=0$~\cite{Raas2009},
\begin{equation}
\int_0^\infty \mr{d}\omega \, \chi_c (\omega) = \la (\delta \hat{n}_i)^2 \ra = \la P_{i0} \ra + \la P_{i2} \ra
  = 1 - \la P_{i1} \ra .
\end{equation}
Indeed, $\chi_c (\omega)$
has a peak largely overlapping with the upper Hubbard
band (e.g. see Fig.~1 in \Ref{Lee2017}).
As a sensible choice for this mean-field parameter,
in practice, we simply take the (exact) numerical data
for $\la P_{i1} \ra$ from our DMFT calculations
(see Fig.~S2 in \Ref{Lee2017}).

Similarly, the second parameter in our mean-field decoupling scheme,
$\Delta_{dh} = \tfrac{v}{2} \sum_\nu \la d_{i\nu} h_{j\nu} + d_{j\nu} h_{i\nu} \ra$,
is a static expectation value with respect to the
ground state of the low-energy effective Hamiltonian.
Thus the value of $\Delta_{dh}$
can be interpreted as a property of Fermi-liquid quasiparticles.
Furthermore, given the general property 
$A_{dd^\dagger} (\omega = 0) = A_{dh}(\omega = 0)
= 1/2\pi$ which is constant
within the metallic phase for arbitrary $U < U_{c2}$ at $T = 0$,
as also supported by our DMFT results in \Ref{Lee2017},
we take this as an indication that the mobility of the
quasiparticles is rather independent of $U$
[similarly, the transmission probability in impurity models
at $T=0$ is related to the value of the spectral function
$A(\omega)$ at $\omega=0$~\cite{Meir1992},
and not the width of the quasiparticle peak around $\omega=0$.]
Accordingly we choose
a constant $\Delta_{dh} = \tfrac{v}{2} \sum_\nu \la d_{i\nu} h_{j\nu} + d_{j\nu} h_{i\nu} \ra$ for $U < U_{c2}$.
We tested different values of $\Delta_{dh}$ as a free parameter,
and found that with $\Delta_{dh} = U_{c2}$ we can reproduce the DMFT
results in \Ref{Lee2017}
of the subpeak position $\omega_p$ over a range of $U$
up to a constant prefactor;
$\omega_p \simeq 4.703 \times \omega_{dh}$.
Here $U_{c2}$ is the critical interaction strength for the metal-to-insulator transition at zero temperature
which our DMFT calculation~\cite{Lee2017} identifies as $U_{c2} = 2.91(1) D$ for the
semi-circular density of states of the lattice.

With the choice of $\Delta_{dh}$ and $\la P_{i1} \ra$ as discussed above,
we demonstrated in \Ref{Lee2017} that
the peak position $\omega_{dh}$ of the analytically
calculated correlation functions
decreases linearly with increasing $U$, with overall qualitative agreement with the DMFT data
[see Fig.~3(b) of \Ref{Lee2017} for details].

\ytableausetup{smalltableaux} 
\begin{table*}
\begin{tabular}{| c | c | c | c | c |}
\hline
$N$ & $\bar{n}$ & $P_{i\bar{n}} P_{j\bar{n}}$ & $P_{i,\bar{n} \pm 1} P_{j,\bar{n} \mp 1}$ & 
$H_{ss}$, $H_{dh}$ \\ \hline
\multirow{2}{*}{\, 2 \,} & \multirow{2}{*}{\, 1 \,} & 
\, $\ydiagram{1} \otimes \ydiagram{1} = 
   \ydiagram{2} \oplus ( \cdot )$ \, &
\, $\ydiagram{1,1} \otimes ( \cdot ) 
   = ( \cdot )$ \, &
\, $( \cdot )$ \, \\
 & & $2 \otimes 2 = 3 \oplus 1$ & $1 \otimes 1 = 1$ & $1$ \\ \hline
\multirow{2}{*}{\, 3 \,} & \multirow{2}{*}{\, 1 \,} & 
\, $\ydiagram{1} \otimes \ydiagram{1} = \ydiagram{2} \oplus \ydiagram{1,1}$ \, &
\, $\ydiagram{1,1} \otimes ( \cdot ) = \ydiagram{1,1}$ \, &
\, $\ydiagram{1,1}$ \, \\
 & & $3 \otimes 3 = 6 \oplus \overline{3}$ & $\overline{3} \otimes 1 = \overline{3}$ & $\overline{3}$ \\ \hline
\multirow{2}{*}{\, 4 \,} & \multirow{2}{*}{\, 2 \,} & 
\, $\ydiagram{1,1} \otimes \ydiagram{1,1} = \ydiagram{2,2} \oplus
   \ydiagram{2,1,1} \oplus ( \cdot ) 
   $ \, &
\, $\ydiagram{1,1,1} \otimes \ydiagram{1} = \ydiagram{2,1,1} \oplus
   ( \cdot ) 
   $ \, &
\, $\ydiagram{2,1,1} \oplus ( \cdot )
$ \, \\
 & & $6 \otimes 6 = 20 \oplus 15 \oplus 1$ & $\overline{4} \otimes 4 = 15 \oplus 1$ & $15 \oplus 1$ \\ \hline
\end{tabular}
\caption{The Young tableaux describing the $\mr{SU}(N)_\mr{flavor}$
symmetry sectors and their degeneracies
for the projectors and the two-site interactions.
Here we use the degeneracy, i.e., dimension, of a multiplet
as an additional label to the Young tabelau, where
$\bar{q}$ refers to the dual representation of $q$.
Given a single local flavor index $\nu=1,\ldots,N$
of fermionic character, the local multiplet for a single lattice site necessarily
needs to be antisymmetric, i.e., a single column  in 
a Young tableau.
The third column for $P_{i\bar{n}} P_{j\bar{n}}$
then combines two such Young tableaux at
the specific integer
filling $\bar{n} \equiv \la \hat{n}_i \ra = \lfloor N/2 \rfloor$
chosen here as well as
in the DMFT calculations in \Ref{Lee2017}.
On the l.h.s.\ of the equations in the fourth column,
one particle (box) has been transferred across the two sites,
as compared with the l.h.s.\ tableaus in the third column.
The last column shows the intersection of the resulting
symmetry sectors of the previous two columns, which thus
represents the relevant block-diagonal
symmetry sectors of $H_{ss}$ and $H_{dh}$.
This is identical with the result of the fourth column for
$P_{i,\bar{n} \pm 1} P_{j,\bar{n} \mp 1}$.
A singlet is given by $( \cdot )$, i.e., no box in the tableau,
or, equivalently, by a full column of $N$ boxes.}
\label{tab:Sym}
\end{table*}

\section{Doublon-holon interaction with more flavors}
\label{sec:SUN}

In \Ref{Lee2017} we have also presented DMFT results for $N > 2$ flavors.
The analysis in \Sec{sec:DHcorrel} cannot predict how the spectral weight of the subpeaks changes 
by considering $N>2$ flavors,
since the analysis above does not consider
the quasiparticle peak and the Hubbard bands at all;
therefore the relative transfer of spectral weight
among different spectral features is beyond the scope of that analysis.
Nevertheless, the low-energy effective interaction $H_{dh}$ in \Eq{eq:Hdh}
does exhibit 
an enlarged degeneracy of the doublon-holon pair excitations with increasing $N$, which originates from the larger $\mr{SU}(N)$ flavor symmetry.
As shown in Fig.~4 in \Ref{Lee2017},
this eventually results in
a wider peak at $\omega=0$ due to the spin
dynamics via $H_{ss}$, as well as
more pronounced subpeaks on
the inner edge of the Hubbard side bands
due to doublon-holon dynamics via $H_{dh}$.

For a more detailed analysis of relative
degeneracies,
we study the $\mr{SU}(N)$ flavor symmetry properties of the two-site interaction terms $H_{ss}$ and $H_{dh}$ in
Eqs.~\eqref{eq:Hss} and \eqref{eq:Hdh}, respectively,
by using Young tableaux.
The building blocks of $H_{ss}$ and $H_{dh}$ are the projected hoppings
$\Pi_{ij}^{n m} = (P_{in} P_{j,n-m-1}) (\sum_\nu c_{i\nu}^\dagger c_{j\nu} ) (P_{i,n-1} P_{j,n-m})$
[cf.~\Eq{eq:Pij}].
In the joint Hilbert space of sites $i$ and $j$,
$\Pi_{ij}^{n m}$ is block diagonal in
the symmetry sectors of the $\mr{U}(1)_\mr{charge} \otimes \mr{SU}(N)_\mr{flavor}$ symmetry, since 
each of the terms above respects this symmetry.

When $U$ is large, 
charge configurations far away from the average occupation $\bar{n}$
are suppressed albeit still present even if
the system is in the metallic phase.
Then the relevant charge sectors
are restricted to $(n_i, n_j) = (\bar{n}, \bar{n})$
for $H_{ss}$ and $(n_i, n_j) = (\bar{n}+1, \bar{n}-1), (\bar{n}-1, \bar{n}+1)$
for $H_{dh}$.
These charge configurations are connected by
the elementary building block of $H_{ss}$ and $H_{dh}$,
namely $\Pi_{ij}^{n m}$ and $(\Pi_{ij}^{n m})^\dagger$.
Thus we will compute the symmetry sectors of the projectors
$P_{i\bar{n}} P_{j\bar{n}}$ and $P_{i,\bar{n}\pm 1} P_{j,\bar{n}\mp 1}$
which act on the left or right of $\Pi_{ij}^{n m}$,
depending on the value of $m$.
In this section, we focus on the filling $\bar{n} = \la \hat{n}_i \ra = \lfloor N/2 \rfloor$ as considered in \Ref{Lee2017}.
A generalization to arbitrary integer filling $\bar{n}$ is straightforward.

\ytableausetup{boxsize=.35em,aligntableaux=center}%

Table~\ref{tab:Sym} shows the $\mr{SU}(N)_\mr{flavor}$ symmetry labels
and the corresponding degeneracies of the projectors and the two-site
interactions. Note that the $\mr{U}(1)_\mr{charge}$ symmetry labels
are trivially the sum $n_i + n_j$.
We observe that the $\mr{SU}(N)_\text{flavor}$ symmetry sectors 
for $P_{i\bar{n}} P_{j\bar{n}}$ and $P_{i,\bar{n}\pm 1} P_{j,\bar{n}\mp 1}$
are different, by comparing the third and the fourth columns in Table \ref{tab:Sym}.
Therefore the symmetry sectors that are relevant
for $H_{ss}$ and $H_{dh}$,
are given by the common sectors, i.e., the intersection
between the sectors for
$P_{i\bar{n}} P_{j\bar{n}}$ and $P_{i,\bar{n}\pm 1} P_{j,\bar{n}\mp 1}$; 
see the last column in Table \ref{tab:Sym}.

For $N = 2$, we notice that the pair states onto which
$H_{ss}$ and $H_{dh}$ project [cf.~\Eq{eq:Hss_Hdh_SU2}]
are in the singlet sector of the $\mr{SU}(2)_\mr{flavor}$ symmetry, 
that is, the spin-spin interaction prefers the spin singlet
(with binding energy $-4 v^2 / U$) and the doublon-holon pair
(once it exists, with excitation energy $4v^2 / U$)
is non-degenerate.
On the other hand,
the low-energy sectors for SU$(N > 2)$ have larger degeneracy,
i.e., lie in symmetry sectors with larger multiplet dimensions.
For $N = 3$, the doublon-holon pair excitation with energy
$4v^2 / U$ and the flavor-flavor
bound state with energy $-4v^2 / U$
are in the 3-fold degenerate sector $\ydiagram{1,1}$,
as seen from the last column in \Tbl{tab:Sym}.
For $N = 4$, the pair excitations have two different energies,
$4v^2 / U$ and $12 v^2 / U$, in the 15-fold degenerate sector
$\ydiagram{2,1,1}$ and the non-degenerate singlet sector
$(\cdot)$, respectively [see the last column in \Tbl{tab:Sym}].
The flavor-flavor
bound states are also in these two sectors
with the respective binding energies $-4v^2 / U$ and $-12 v^2 / U$.

This symmetry argument explains
the strong enhancement of the subpeaks 
observed \cite{Lee2017}
in the local spectral functions $A(\omega)$
for larger $N$ 
(cf.~Fig.~4 in \Ref{Lee2017}),
compared with $N = 2$ case 
(cf.~Fig.~1 in \Ref{Lee2017}).
The subpeaks gain more spectral weight for larger $N$ and even become higher than the rest of the Hubbard bands for $N = 4$, consistent with the increasing degeneracy of doublon-holon pairs: $1$ for $N = 2$, $3$ for $N = 3$, and $15 \oplus 1$ for $N = 4$. On the other hand, the quasiparticle peak around $\omega = 0$ is also enhanced for larger $N$; the quasiparticle peak persists even at elevated $U$, supported by the degeneracy in the flavor-flavor terms that grows in the same way as the degeneracy of the doublon-holon terms.
Overall then, by the sum rule conservation of local
correlations functions, accordingly,
the Hubbard bands have lower relative weight and height.

\section{Conclusion}
\label{sec:conclusion}

We have used a generalized SWT
to obtain an effective low-energy Hamiltonian for multi-flavor Hubbard models
which contains all the effective interactions up to
order $O(1/U)$.
Our straightforward approach avoids
the need of determining the appropriate canonical
transformation as required in previous SWT schemes~\cite{Harris1967,*Chao1977,*MacDonald1988,*Eskes1994,*Eskes1994a},
e.g., used for the derivation of the $t$-$J$ model.

Having derived the effective Hamiltonian,
we interpreted the Fermi-liquid ground state of the
paramagnetic metallic phase as
the condensate of the doublon-holon pairs, 
and introduced a mean-field variable based on the expectation value of the doublon-holon pair annihilation operator.
This mean-field approximation is analogous to the Bardeen-Cooper-Schrieffer (BCS) theory of superconductivity
in which the expectation value of the Cooper pair annihilation operator is chosen as the superconducting order parameter.
Thus the role of the doublon-holon interaction term is crucial here,
reminiscent of the Cooper pair terms in the BCS Hamiltonian.

Then we computed the correlation functions of doublons and holons, focusing on the well-separated
intermediate energy scale in between 
the low- and high-energy scales associated with the features in the local spectral function, namely the smaller scale for the quasiparticle peak
and the larger scale for the Hubbard bands. Since
these features are associated with the spin and charge degrees of freedom,
respectively, we effectively used a mean-field decoupling of
the spin and charge degrees of freedom from doublons and holons.

We observed subpeaks at finite frequency that
are clearly associated with the doublon-holon dynamics.
The numerical results shown in \Ref{Lee2017} and the symmetry argument in \Sec{sec:SUN} 
demonstrate
that the subpeaks become more pronounced for larger number $N$ of particle flavors,
since the doublon-holon excitation on a pair of nearest-neighbour sites gains access to a larger degenerate state space.

We expect that the subpeaks would be observable in photoemission
  spectroscopy experiments of correlated materials.  The subpeaks in
  the local spectral function correspond to dispersive features in the
  momentum-resolved spectral function which are distinguishable from
  those for the quasiparticle peak and the Hubbard bands; see
  \Ref{Lee2017} for details.  The enhancement of the subpeaks for
  larger $N$ is relevant to multi-band materials. However, for
  materials in which the Hund's
  coupling~\cite{deMedici2011,Stadler2015} is important, further
  analysis (beyond the scope of this work) would be needed to
  investigate how it affects the doublon-holon dynamics discussed
  here.  Another promising class of systems for probing the effects of
  doublon-holon dynamics would be the cold atom systems studied in
  Refs.~\cite{Taie2012,Hofrichter2016}, where the $\mr{SU}(N)$ Hubbard
  model has been realized with exact $\mr{SU}(N)$ symmetry and tunable
  values of $N$.

\begin{acknowledgments}
We thank M. Bukov, K. Penc, A. Polkovnikov, and M. Punk for fruitful discussion.
This work was supported by Nanosystems Initiative Munich.
S.-S.B.L. acknowledges support from the Alexander von Humboldt Foundation and the Carl Friedrich von Siemens Foundation,
A.W. from the German Research Foundation (DFG) WE4819/2-1
\end{acknowledgments}

%

\end{document}